\documentclass[prb,showpacs,twocolumn,aps,epsf]{revtex4}
\usepackage{feynmp}
\usepackage{amssymb}
\usepackage{epsfig}
\usepackage{graphicx}
\usepackage{subfigure}
\usepackage{color}

\begin{document}

\title{Paring instability in the mixed state of $d$-wave superconductor}
\author{Hua Jiang, Guo-Zhu Liu and Geng Cheng \\
%EndAName
{\small {\it Department of Modern Physics, University of Science
and Technology of China, Hefei, Anhui, 230026, P.R. China }}}

\begin{abstract}
We propose that an excitonic gap can be generated along nodal
directions by Coulomb interaction in the mixed state of $d$-wave
cuprate superconductors. In a superconductor, the Coulomb
interaction usually can not generate any fermion gap since its
strength is weakened by superfluidity. It becomes stronger as
superfluid density is suppressed by external magnetic field, and
is able to generate a gap for initially gapless nodal
quasiparticles beyond some critical field $H_{c}$. By solving the
gap equation, it is found that the nodal gap increases with
growing field $H$, which leads to a suppression of thermal
conductivity at zero temperature. This mechanism naturally
produces the field-induced thermal metal-insulator transition
observed in transport experiments.
\end{abstract}


\maketitle

%%%%%%%%%%%%%%%%%%%%%%%%%%%%%Main Body%%%%%%%%%%%%%%%%%%%%%%%%%%%%%%%%%%%%%

\section{Introduction}

The low-energy spectral and transport properties of nodal
quasiparticles in $d$-wave cuprate superconductor are very
important issues. In the absence of external magnetic field, the
ground state is occupied by uniform superconductivity. For a clean
superconductor, the low-energy density of states vanishes linearly
as $N(\omega) \sim |\omega|$ upon approaching the Fermi surface.
It acquires a finite value at zero-energy in the presence of weak
impurity scattering. In this case, the nodal quasiparticles
exhibit universal transport behavior in the sense that the
electric, thermal and spin conductivities are independent of
impurity scattering rate at the $\omega \rightarrow 0, T
\rightarrow 0$ limit \cite{Lee93, Durst, Hussey}. The zero
temperature thermal conductivity $\kappa$ is of particular
interests since it is not affected by vertex corrections
\cite{Durst}. Remarkably, the predicted universal thermal
conductivity has been confirmed by heat transport measurements at
optimal doping \cite{Taillefer, Chiao}.

When placed in an external perpendicular magnetic field, the
superconductor enters into the mixed state in the range $H_{c1} <
H < H_{c2}$. Inside the vortex cores, the superfluid current is
significantly reduced by the magnetic field. The low-energy
fermionic excitations in the mixed state are expected to have
rather different low-energy behaviors comparing with those in the
uniform zero-field condensate \cite{Hussey}. As revealed by heat
transport measurements, the thermal conductivity loses its
universality and depends on the impurity scattering rate
\cite{Hussey}. In addition, on the underdoping side, it decreases
as the magnetic field grows up \cite{Hawthorn, Sun03}. There seems
to be a field-induced thermal metal-to-insulator transition in
some underdoped cuprate superconductors \cite{Hawthorn, Sun03}.
These experimental results can be intuitively understood by
assuming that the nodal fermions acquire finite mass gap in the
mixed state. A phenomenological expression for the nodal gap was
proposed \cite{Gusynin} to understand the field-induced reduction
of thermal conductivity. However, the dynamic origin for the gap
generation has not been discussed.

The goal of this paper is to suggest a mechanism for opening the
field-induced gap for the initially gapless nodal quasiparticles.
Generally, this mechanism would be realized by some kind of
fermion self-interaction or boson mediated interaction. Such
interaction should have the following two features: it is weak
enough to be irrelevant in the uniform superconducting state; it
gets stronger with growing perpendicular magnetic field so that a
finite gap is generated beyond some critical magnetic field
$H_{c}$.

Qualitatively, the U(1) gauge fluctuation arising from strong
correlation provides a good candidate for such mechanism. The
\emph{t}-\emph{J} model of cuprate superconductors can be
theoretically treated by the slave boson method. After making
mean-field analysis and including fluctuations, there appears an
emergent U(1) gauge field which interacts strongly with
spin-carrying spinons and charge-carrying holons \cite{Leermp}.
The superconductivity is realized by holon condensation below
$T_{c}$, while the $d$-wave energy gap is formed by spinon
pairing. In the superconducting state, the low-energy elementary
excitations are gapless nodal spinons and the U(1) gauge boson is
gapped via the Anderson-Higgs mechanism. The finite gauge boson
gap weakens gauge interaction, so usually no fermion gap can be
generated. However, once the superfluid density is suppressed by
external magnetic field, the gap of U(1) gauge boson decreases and
the strength of gauge interaction increases with growing magnetic
field \cite{Liu03}. Then a finite gap for nodal spinons could be
generated by gauge fluctuation, leading to suppression of thermal
conductivity. Unfortunately, it is hard to average over the vortex
distributions within this formalism due to the complexity brought
by spin-charge separation.

The gapless nodal fermions might acquire a gap via the magnetic
catalyst mechanism \cite{Gusynin94} when they are placed in an
external magnetic field. But this mechanism depends on a crucial
assumption that the fermion stays in the lowest Landau level
\cite{Gusynin94, Ferrer}. However, in the case of high temperature
superconductor, the Landau level has been shown not to be the
appropriate description of fermion energy spectrum in the mixed
state \cite{Franz99}. Therefore, the magnetic catalyst mechanism
is unlikely to be at work.

In this paper, we study the possibility of gap generation due to
the long-range Coulomb interaction between charged nodal
quasiparticles. Two quasiparticles that carry the same charges
always experience a repulsive Coulomb force, while the
quasiparticle and quasihole experience an attractive Coulomb
force. When the attractive force is sufficiently strong, it is
possible that a Dirac quasiparticle is combined with a Dirac
quasihole to form a stable excitonic pair. Through this mechanism,
the gapless fermion acquires a finite excitonic gap.

Recently, this kind of gap generation was argued to lead to an
insulating ground state in single layer graphene, when the Coulomb
interaction strength $g$ is larger than some threshold $g_{c}$ and
the fermion flavor $N$ is less some threshold $N_{c}$
\cite{Khveshchenko, Gorbar, Liu08}. Moreover, an interesting
superfluidity was predicted to exist in bi-layer graphene based on
a similar paring instability \cite{MacDonald}. The long-range
Coulomb interaction is also very important in cuprate
superconductors. First of all, it lifts the gapless Goldstone mode
up to plasmon mode, which is actually the rudiment of Higgs
mechanism. Its importance in the formation of stripe phase has
been emphasized by several authors \cite{Kivelson}. However, its
role and influence on nodal quasiparticles are still in debate
\cite{Emery}. In the absence of a reliable microscopic theory of
Coulomb interaction, we resort to the phenomenological approach.

From the available extensive experiments, we know that the nodal
quasiparticles have rather long mean free path and behave like
well-defined Bogoliubov-Landau quasiparticles in the uniform
superconducting state \cite{Orenstein, Housseini}. This fact and its
excellent agreement with BCS-type analysis \cite{Lee93, Durst,
Orenstein} implies that the Coulomb interaction must be fairly weak
in the superconducting state and generally can not generate any
fermion gap, except in the lightly doping region. On the other hand,
in the non-superconducting ground state, it is generally believed
that there are no well-defined Landau quasiparticles. It is
reasonable to expect that the long-range Coulomb interaction is very
strong in this state. The field-induced mixed state lies between
these two extreme limiting cases. As the superfluid density
decreases with magnetic field, the effective strength of Coulomb
interaction gets stronger. For sufficiently strong interaction, a
dynamical fermion gap can be generated by forming excitonic pairs.
To implement this intuitive picture with ( explicit ) computations,
we assume a phenomenological form for the effective interaction
strength which is a function of magnetic field $H$. After solving
the associated gap equation, we find that the growing magnetic field
drives the system towards a phase transition into excitonic
insulating state beyond some critical value $H_{c}$. Once the nodal
fermion acquires a finite gap $m$, the low-energy fermionic
excitations are significantly suppressed below the scale $m$,
leading to reduction of thermal conductivity. ( This can help to
understand the transport behaviors observed in the mixed state of
cuprate superconductors. )

Besides the thermal metal-insulator transition, another important
issue about field-induced phenomena is the enhancement of
antiferromagnetic correlations inside the vortex cores. Such
microscopic coexistence of magnetic order and superconductivity has
been investigated experimentally \cite{Lake01, Lake02, Hoffman,
Mitrovic, Khaykovich, Khaykovich05, Chang} and theoretically
\cite{Demler, Kivelson02}. From the field theoretic point of view,
the field-induced antiferromagnetism or spin density wave can be
represented by a mass term for nodal fermions \cite{Gusynin,
Tesanovic02}. With this identification, the mechanism responsible
for the thermal metal-insulator transition can also account for the
existence of field-induced magnetic order in the mixed state.

Before presenting the technical details, we would like to point out
that a number of assumptions and approximations will be used to
simplify discussions on the issue of dynamical gap generation. Thus,
the conclusions reached in this paper are reliable only at the
qualitative, rather than quantitative, level.

The paper is arranged as follows. In Sec. II, we build the model
and write down the gap equation. In Sec. III, we propose the
phenomenological form of the effective Coulomb interaction and
calculate its dependence on magnetic field $H$. In Sec. IV, we
solve the gap equation and give the field dependence of critical
coupling $g_{c}$ and dynamical gap. The qualitative understanding
of transport experiments is also discussed. We ends with a summary
and discussion in Sec. V.

\section{Model Hamiltonian and gap equation}

We begin our discussion with the following Hamiltonian of $d$-wave
superconductor
\begin{equation}
H_{0} = \sum_{\mathbf{k}} \Phi_{\mathbf{k}}^{\dagger}
[\epsilon_{\mathbf{k}}\tau_{3} - \Delta_{\mathbf{k}}\tau_{1}]
\Phi_{\mathbf{k}},
\end{equation}
where the standard two-component Nambu spinor representation
$\Phi_{\mathbf{k}}^{\dagger} = (c_{\mathbf{k}\uparrow}^{+},
c_{-\mathbf{k}\downarrow})$ is adopted and $\tau_{i}$ is Pauli
matrix. The electron dispersion is $\epsilon_{\mathbf{k}} =
-2t(\cos k_{x}a + \cos k_{y}a) - \mu_{0}$ with $\mu_{0}$ being the
chemical potential and the $d$-wave energy gap is
$\Delta_{\mathbf{k}} = \frac{\Delta_{0}}{2} (\cos k_{x}a - \cos
k_{y}a)$. The quasiparticle spectrum is $E_{\mathbf{k}} =
\sqrt{\epsilon_{\mathbf{k}}^{2} + \Delta_{\mathbf{k}}^{2}}$, which
has four nodal points at the Fermi level. Linearizing the
dispersion in the vicinity of the nodes, one obtains the spectrum
$E_{\mathbf{k}} = \sqrt{v_{F}^{2}k_{1}^{2} +
v_{\Delta}^{2}k_{2}^{2}}$, where $\mathbf{k_{1}}$ is perpendicular
to the Fermi surface and $\mathbf{k_{2}}$ is parallel to the Fermi
surface. The four-component Dirac spinor can be defined as
\cite{Herbut02, Gusynin, Sharapov}
\begin{eqnarray}
\Psi_{1(2)}^{\dagger}(\mathbf{q},\omega_{n}) &=&
(c_{\uparrow}^{\dagger}(\mathbf{k},\omega_{n}),
c_{\downarrow}(-\mathbf{k},-\omega_{n}),
\nonumber \\
&&
c_{\uparrow}^{\dagger}(\mathbf{k}-\mathbf{Q}_{1(2)},\omega_{n}),
c_{\downarrow}(-\mathbf{k}+\mathbf{Q}_{1(2)},-\omega_{n})),
\nonumber
\end{eqnarray}
where $\mathbf{Q}_{1(2)}=2\mathbf{K}_{1(2)}$ is the wave vector
that connects the nodes within the diagonal pairs,
$\mathbf{k}=\mathbf{K_{i}}+\mathbf{q}$ with
$\mathbf{q}\ll\mathbf{K_{i}}$. Here, we use the four-component
spinor because it is impossible to define chiral symmetry in
two-component representation of fermion field in (2+1) dimensions.

The continuum Hamiltonian of free Dirac fermions can be written as
\begin{eqnarray}
H_{0} = i\int d^{2}\mathbf{r}
\bar{\Psi}_{1}(\gamma_{1}v_{F}\partial_{x} +
\gamma_{2}v_{\Delta}\partial_{y})\Psi_{1} + (1\rightarrow
2,x\leftrightarrow y),
\end{eqnarray}
where $\bar{\Psi} = \Psi^{\dagger}\gamma_{0}$. The $4\times 4$
matrices can be chosen as $\gamma_{0}=\sigma_{1}\otimes \tau_{0}$,
$\gamma_{1}=-i\sigma_{2}\otimes \tau_{3}$, and
$\gamma_{2}=i\sigma_{2}\otimes \tau_{1}$, where $\sigma_{i}$ acts in
the subspace of the nodes in a diagonal pair, $\tau_{i}$ acts on
indices inside a Nambu field. There are two matrices anticommuting
with them, $\gamma_{3}=i\sigma_{2}\otimes \tau_{2}$, and
$\gamma_{5}=\sigma_{3}\otimes \tau_{0}$. The matrices satisfy the
Dirac algebra $\{\gamma_{\mu}, \gamma_{\nu}\} = $2diag(1,-1,-1).

The Hamiltonian for the Coulomb interaction is
\begin{equation}
H_{C}=\frac{1}{4\pi}\sum_{i,i^{\prime}=1}^{N}\int_{\mathbf{r},\mathbf{r}^{\prime}}
\bar{\Psi}_{i}(\mathbf{r})\gamma_{0}\Psi_{i}(\mathbf{r})\frac{g}{|\mathbf{r}-\mathbf{r}^{\prime}|}
\bar{\Psi}_{i^{\prime}}(\mathbf{r}^{\prime})\gamma_{0}\Psi_{i^{\prime}}(\mathbf{r}^{\prime}),
\end{equation}
The bare Coulomb interaction in momentum space is $V_{0}(q) =
g/|\mathbf{q}|$. The parameter $g$ measures the strength of bare
Coulomb interaction in the non-superconducting ground state
\cite{note1}. It is easy to see that this model is very similar to
that in single layer graphene, where the low-energy excitations are
also massless Dirac fermions. Unlike the semi-metal background in
graphene, in the present case the Dirac fermions are in a charge
condensate, which unavoidably affect their properties and the
Coulomb interaction between them.

From the results obtained in graphene, it is known that sufficiently
strong Coulomb interaction can lead to finite excitonic gap, when
the fermion flavor is below a critical value \cite{Khveshchenko,
Gorbar, Liu08}. We speculate that a similar pairing instability to
occur in the mixed state of $d$-wave superconductor. When studying
the gap generation, we fix the physical fermion flavor $N=2$. Thus,
coupling $g$ becomes the only variable that tunes the excitonic
phase transition.

The Hamiltonian is invariant under the continuous chiral
transformation $\Psi \rightarrow e^{i\theta \gamma_{3(5)}} \Psi$. It
will be dynamically broken once the Dirac fermion acquires a finite
mass via the effective Coulomb interaction. This phenomenon is
non-perturbative in nature and generally can be studied by analyzing
the self-consistent Dyson equation
\begin{equation}
G^{-1}(p) = G_{0}^{-1}(p) +
T\sum_{n=-\infty}^{\infty}\int\frac{d^{2}\mathbf{k}}
{(2\pi)^{2}}\gamma_{0}G(k)\Gamma_{0}(p,k) V(p-k),
\end{equation}
where the (2+1)-dimensional momentum is defined as
\begin{equation}
k=(i\omega_n,\mathbf{k}).
\end{equation}
The Matsubara frequency is $\omega_n=(2n+1)\pi T$ for fermions and
$\omega_n = 2n\pi T$ for bosons. Here, $\Gamma_{0}(p,k)$ is the full
vertex function. The free propagator for massless Dirac fermion is
\begin{equation}
G_{0}(k) =
\frac{1}{i\omega_n\gamma_{0}-v_{F}k_{1}\gamma_{1}-v_{\Delta}k_{2}\gamma_{2}}.
\end{equation}
Due to the Coulomb interaction, it becomes the complete propagator
$G(p)$, which is determined by the Dyson equation. In the case of
QED$_{3}$, the dynamical chiral symmetry breaking can be most
conveniently studied using the $1/N$ expansion \cite{QED3}. Here we
follow the same strategy and keep only the leading order of $1/N$
expansion. So we can neglect the wave function renormalization and
replace the vertex function $\Gamma_{0}$ by $\gamma_{0}$. Now the
complete propagator can be formally written as
\begin{equation}
G(k) =
\frac{1}{i\omega_n\gamma_{0}-v_{F}k_{1}\gamma_{1}-v_{\Delta}k_{2}\gamma_{2}
- m(k)},
\end{equation}
where $m(k)$ denotes the Dirac fermion mass. Further, as shown in
the context of QED$_{3}$, at least at low energies and to the
leading order of $1/N$ expansion, the velocity anisotropy is
irrelevant to the critical behavior \cite{Vafek}. We simply set
$v_{F}=v_{\Delta}=1$ whenever they multiply the momenta
($k_{1},k_{2}$) in the gap equation.

The full Coulomb interaction function is
\begin{equation}
V(q) = \frac{1}{|\mathbf{q}|/g  + N\chi(q)}.
\end{equation}
The polarization function $\chi(q)$ contains all information about
how the Dirac fermions response to the many-particle system. We
first consider the non-superconducting ground state. Within the
random phase approximation, the fermion propagator $G$ and the
vertex function $\Gamma_{0}$ are both replaced by the bare ones,
i.e.,
\begin{equation}
\chi(q) = -T\sum_{n=-\infty}^{\infty}
\int\frac{d^{2}\mathbf{k}}{(2\pi)^{2}}
\mathrm{Tr}[\gamma_{0}G_{0}(k)\gamma_{0}G_{0}(k-q)].
\end{equation}
Inserting the expression for the interaction function, the gap
equation can now be written as
\begin{equation}
m(p)= T \sum_{n=-\infty}^{\infty}\int
\frac{\mathrm{d}^2\mathbf{k}}{(2\pi)^2}
\frac{m(k)}{\omega_{n}^{2}+\mathbf{k}^{2}+m^2(k)}
\frac{1}{|\mathbf{q}|/g + N\chi(q)},
\end{equation}
where $q=p-k$. In the instantaneous approximation, $\chi(q)$ has the
following zero frequency expression \cite{Dorey}
\begin{equation}
\chi(\mathbf{q})=\frac{2T}{\pi}\int_{0}^{1}\mathrm{d}x
\log{[2\cosh{\frac{\sqrt{x(1-x)}|\mathbf{q}|}{2T}}]}.
\end{equation}
Now the gap $m$ is independent of frequency and the frequency
summation can be carried out with the result
\begin{equation}
m(T,\mathbf{p})= \int\frac{\mathrm{d}^2\mathbf{k}}{8\pi^2}
\frac{m(T,\mathbf{k})}{\sqrt{\mathbf{k}^{2} + m^2(T,\mathbf{k})}}
\frac{\tanh{\frac{\sqrt{\mathbf{k}^{2}+m^2(T,\mathbf{k})}}{2T}}}
{|\mathbf{q}|/g + N\chi(\mathbf{q})}.
\end{equation}
In the limit of zero temperature, $\chi(\mathbf{q})=|\mathbf{q}|/8$,
the gap equation further simplifies to
\begin{equation}
m(\mathbf{p})= \int\frac{\mathrm{d}^2\mathbf{k}}{8\pi^2}
\frac{m(\mathbf{k})}{\sqrt{\mathbf{k}^{2} + m^2(\mathbf{k})}}
\frac{1}{|\mathbf{q}|/g+N|\mathbf{q}|/8}.
\end{equation}
The nontrivial solution $m(\mathbf{p})$ of this integral equation
signals the occurrence of dynamical mass generation.

When the ground state is occupied by the uniform
superconductivity, the Coulomb interaction function must be
modified. In the mixed state, the superfluid density is a
non-uniform quantity which has different values at different
spatial positions. To study the gap equation in the mixed state,
we should average over the vortices and obtain a mean value of
superfluid density. This is the task of the next section.

\section{Ansatz of effective interaction}

In the underdoping and optimal doping regions, the ground state is
occupied by the superconductivity, which significantly weakens the
Coulomb interaction between Dirac fermions. Such effect can be
described by calculating the polarization function that incorporates
the effect of finite superfluid density. However, it is not clear
how to correctly calculate the polarization function in the
superconducting state, so we will assume a phenomenological form for
the effective interaction function. From the experimental facts, we
know that the interaction must reaches its minimal value when the
superfluid density $\Lambda_{s}$ takes its maximal value. As
$\Lambda_{s}$ decreases with growing magnetic field $H$, the
effective interaction strength increases and eventually takes its
maximal value after the superconductivity is completely destroyed.
We assume the following \emph{ansatz} for the effective interaction
function in the superconducting state
\begin{equation}
V(q, H) = \frac{1}{|\mathbf{q}|/g +
N\chi(q)}\frac{1}{1+\alpha\Lambda_{s}(H)},
\end{equation}
where $\alpha$ is an adjustable parameter. This is the simplest
function that can describe the reduction of strength by superfluid
density. It is largest at the limit $\Lambda_{s} = 0$, and is
smallest at the limit $H=0$. The function $(1+
NV_{0}(q)\chi(q))(1+\alpha\Lambda_{s}(H))$ can be considered as the
effective dielectric function of superconducting state. In the mixed
state, the field-dependent superfluid density $\Lambda_{s}(H)$
controls the effective strength of Coulomb interaction. The
parameter $\alpha$ must be properly chosen so that a moderately
strong magnetic field $H_{c}$ separates the gapless and gaped
phases. In order to see how the critical point depends on $H$, we
need to solve the gap equations after including $\Lambda_{s}(H)$.

To study the gap equation, the superfluid density $\Lambda_{s}(H)$
should be obtained by averaging over the vortices. In the mixed
state, the low-energy properties of $d$-wave superconductor are
dominated by the extended quasiparticles in the bulk material,
unlike the case of conventional $s$-wave superconductor. Volovik
\cite{Volovik} proposed a semiclassical approach and showed that
the density of states varies as $\sqrt{H}$ at low temperatures,
which has been observed by experiments \cite{Hussey}. Within the
semiclassical treatment, the effects of circulating supercurrent
around vortices can be represented by a Doppler shift \cite{Yu,
Kubert, Vekhter} in the quasipartilce spectrum, $\omega
\rightarrow \omega + \mathbf{k} \cdot \mathbf{v}_{s}(\mathbf{r})$,
where $\mathbf{v}_s(\mathbf{r})$ is the superfluid velocity at a
position $\mathbf{r}$ and $\mathbf{k}$ is the quasiparticle
momentum which can be approximated by its value at the node. Then
the fermion Green function can be written as
$G(\omega,\mathbf{k},\mathbf{r}) = G(\omega +
\epsilon_i(\mathbf{r}),\mathbf{k})$, where
$\epsilon_i(\mathbf{r})= \mathbf{k}_i \cdot
\mathbf{v}_{s}(\mathbf{r})$. The local value $F(\mathbf{r})$ of
any physical quantity $F$ determined by the Green function can be
obtained using the above local Green function. The field-dependent
quantity $F(H)$ is written as the following spatial average $F(H)
= \frac{1}{A}\int d^{2}\mathbf{r}F(\omega+\epsilon (\mathbf{r}))$,
where the integral is taken over a unit cell of the vortex lattice
with area $A$. Such averaging integral depends on the vortex
distribution. The field-dependent quantity is
\begin{equation}
F(\omega,H) = \int_{-\infty}^{\infty} d\epsilon F
(\omega+\epsilon) \mathcal{P} (\epsilon),
\end{equation}
with probability function $\mathcal{P} (\epsilon) =
\frac{1}{A}\int d^{2}\mathbf{r}\delta (\epsilon-\mathbf{k}\cdot
\mathbf{v}_{s}(\mathbf{r}))$.

There are several possible choices of $\mathcal{P}(\epsilon)$,
which were discussed in Ref. \cite{Vekhter}. For example, the
distribution function of vortex liquid or solid is
$\mathcal{P}(\epsilon) = \frac{E_{H}^{2}}{2(\epsilon^{2} +
E_{H}^{2})^{3/2}}$; for disordered vortex state, it takes the form
$\mathcal{P}(\epsilon) = \frac{1}{\sqrt{\pi}E_{H}}
\exp\left(-\frac{\epsilon^{2}}{E_{H}^{2}} \right)$. The typical
energy scale of Doppler shift is $E_{H} = \frac{v_{F}}{2R}=
\frac{v_{F}}{2}\sqrt{\frac{\pi H}{\Phi_{0}}}$, where
$R=(\Phi_0/\pi H)^{1/2}$ is the radius of the unit cell of vortex
lattice and $\Phi_{0}=hc/2e$ is one quantum of magnetic flux. The
field dependence of a physical quantity, such as density of state
or specific heat, depend somewhat on the choices of distribution
function, but the qualitative result is not sensitive to the
choice.

The computation of superfluid density $\Lambda_{s}(H)$ within the
semi-classical approximation has already been performed in Ref.
\cite{Sharapov}, so we just list the basic steps and cite the
results. The superfluid stiffness is given by \cite{Durst,
Sharapov}
\begin{equation}
\Lambda_{s}^{ij}(T,H)=\tau^{ij}-\Lambda_{n}^{ij}(T,H),
\end{equation}
where $\tau_{ij}$ is the diamagnetic tensor and $\Lambda_{n}$
represents the normal fluid density divided by the carrier mass. In
the Matsubara formalism, the normal fluid density is \cite{Sharapov}
\begin{eqnarray}
\Lambda_{n}^{ij} &=& -T\sum_{n=-\infty}^{\infty} \int_{\mathrm{HBZ}}
\frac{d^{2}\mathbf{k}}{(2\pi)^{2}}
v_{F}^{i}(\mathbf{k})v_{F}^{j}(\mathbf{k}) \nonumber \\
&& \mathrm{Tr}[G(i\omega_{n},\mathbf{k})\gamma_{0}
\gamma_{5}G(i\omega_{n},\mathbf{k})\gamma_{0}\gamma_{5}],
\end{eqnarray}
with the $i$-component velocity $v_{F}^{i}$. Here, HBZ means the
halved Brillouin zone of the $d$-wave superconductors, i.e., the
domain with two neighboring nodes. The averaged normal fluid density
is given by \cite{Sharapov}
\begin{eqnarray}
\Lambda_{n}^{ij}(H) &=& \int_{-\infty}^{\infty} d\epsilon
\mathcal{P}(\epsilon)\int_{\mathrm{HBZ}}\frac{d^{2}\mathbf{k}}{(2\pi)^{2}}
\int_{-\infty}^{\infty}d\omega \tanh\frac{\omega}{2T}
\frac{v_{F}^{i}v_{F}^{j}}{4i\pi} \nonumber \\
&& \times \mathrm{Tr}[G_{A}(\omega -
\epsilon,\mathbf{k})\gamma_{0}\gamma_{5}G_{A}(\omega -
\epsilon,\mathbf{k})\gamma_{0}\gamma_{5} \nonumber \\
&& - G_{R}(\omega -
\epsilon,\mathbf{k})\gamma_{0}\gamma_{5}G_{R}(\omega -
\epsilon,\mathbf{k})\gamma_{0}\gamma_{5}].
\end{eqnarray}

In principle, the superfluid density should be calculated by
including the complete fermion propagators $G_{R,A}(\omega,
\mathbf{k})$ into (18). Hence it actually satisfies an equation that
couples self-consistently to the gap equation. It is not an easy
task to solve these coupled equations in practice. However, there is
a remarkable simplification if we are mainly interested in what
happens in the vicinity of the critical point of chiral phase
transition. Near the bifurcation point, the gap equation can be
linearized and the gap appearing in the superfluid density can be
taken to be zero. In the limit $m \rightarrow 0$, the normal fluid
density finally becomes \cite{Sharapov}
\begin{equation}
\Lambda_{n}(H) = \frac{v_F}{2\pi v_{\Delta}}
\int_{-\infty}^{\infty}d\epsilon \mathcal{P}(\epsilon)
J(m,\epsilon),
\end{equation}
with $J(\epsilon)=2|\epsilon|$. If we adopt the distribution
function $\mathcal{P}(\epsilon)$ of vortex liquid, then the
superfluid density
\begin{equation}
\Lambda_s(H) = \tau - \frac{v_F}{\pi v_{\Delta}}E_{H}.
\end{equation}
Here, $\tau$ is the zero temperature superfluid density in the
absence of magnetic field and the ratio between $v_{F}$ and
$v_{\Delta}$ appears as a coefficient. Although the anisotropy is
irrelevant when $v_{F(\Delta)}$ multiplies particle momenta, the
ratio might be important in this expression. We simply set the ratio
$v_{F}/v_{\Delta} = 20$ in the following discussions.

\section{Field dependence of critical coupling and thermal
conductivity}

After getting the effective Coulomb interaction, we now study the
self-consistent gap equation
\begin{eqnarray}
m(\mathbf{p})= \int\frac{\mathrm{d}^2\mathbf{k}}{8\pi^2}
\frac{m(\mathbf{k})}{\sqrt{\mathbf{k}^{2} + m^2(\mathbf{k})}}
\frac{1}{|\mathbf{q}|/g+N|\mathbf{q}|/8}\frac{1}{1+\alpha\Lambda_s(H)}.
\nonumber \\
\end{eqnarray}
This integral equation can be solved numerically by the parameter
embedding method, with the Coulomb interaction strength $g$ being
the turning parameter. There is a critical value $g_{c}$ that
separates the chiral symmetric phase ($m=0$ for $g < g_{c}$) from
the symmetry breaking phase ($m \neq 0$ for $g > g_{c}$). $g_{c}$ is
just the critical point of chiral phase transition. We can see that
excitonic pairing is quite different from conventional BCS type
pairing: the former is produced only by sufficiently strong,
attractive Coulomb force between particles and holes, while the
latter is triggered by arbitrary weak attractive force between
electrons. The excitonic gap breaks the chiral symmetry and leads to
the formation of antiferromagnetism or spin density wave in the
field-induced vortex state.

Following Ref. \cite{Sharapov}, the zero temperature superfluid
density is taken to be $\tau=1500\mathrm{K}$ and the energy scale
$E_{H} \sim 30\sqrt{H} \mathrm{K}\mathrm{T}^{-1/2}$. The parameter
$\alpha$ is a variable (in unit of $\mathrm{eV}^{-1}$) depending
on doping concentration and the type of cuprate superconductors.
It surely is not a universal quantity and can not be uniquely
determined. For completeness, we consider a number of possible
values, $\alpha = 4,5,6,7$. For each value of $\alpha$, the
relationship between critical strength $g_{c}$ and magnetic field
$H$ is shown in Fig. 1. As it turns out, critical strength $g_{c}$
decreases as the magnetic field $H$ grows up.

\begin{figure}[ht]
\begin{center}
\epsfxsize=0.40\textwidth \epsfbox{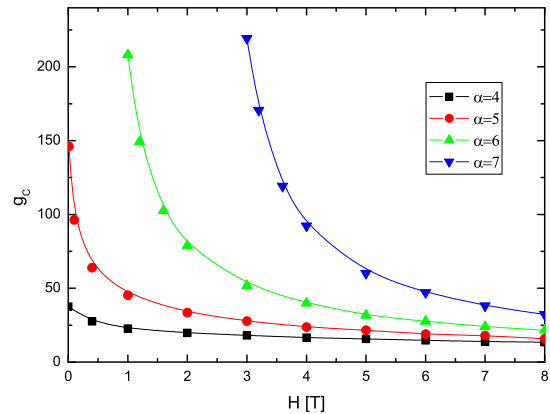}
\end{center}
\caption{The dependence of critical strength $g_{c}$ on magnetic
field $H$ for several choices of $\alpha$.} \label{fig1}
\end{figure}

To judge whether an excitonic gap is generated for nodal fermions,
we can simply compare the physical strength $g$ with the critical
value $g_{c}$. Admittedly the exact value of physical strength $g$
in the non-superconducting ground state is unknown. However, we can
make a simple comparison between its value in cuprate
superconductors with that in single layer graphene. In graphene, the
typical value $g$ is about $\sim 20$. It is not unreasonable to
estimate that the parameter $g$ in the non-superconducting ground
state be larger than $20$, since the correlation is known to be very
strong in cuprate superconductors.

If we assume that $g = 50$, then the critical field $H_{c} = 0, 1.0,
3.5, 6.0$ for parameters $\alpha = 4, 5, 6, 7$ respectively. For
$\alpha = 4$, a finite excitonic gap is generated even in zero field
case. This is able to explain several experimental results performed
in some underdoped cuprates in the absence of external field $H$:
the finite nodal gap found by photoemission \cite{Shenkm}, the
suppression of thermal conductivity $\kappa$ from the universal
value with lowering doping concentration \cite{Sutherland03}, and
the coexistence of competing magnetic order with superconductivity
\cite{Khaykovich05}. For larger values $\alpha = 5, 6, 7$, the gap
is generated only for magnetic field $H$ larger than some critical
value $H_{c}$. These parameters are relevant to the doping regions
in which the nodal gap and competing order appear only in the
field-induced state \cite{Lake01, Lake02, Hoffman, Mitrovic,
Khaykovich}. It appears that the phenomenological parameter $\alpha$
should be an increasing function of doping concentration. The cases
for other values of $g$ and $\alpha$ can be analyzed similarly.

The generated gap will surely affect all observable physical
quantities, such as specific heat, electric and thermal
conductivity. Here we are primarily interested in the zero
temperature thermal conductivity $\kappa$. If we assume a constant
fermion gap $m$ and a small impurity scattering rate
$\Gamma_{\mathrm{imp}}$, then the zero temperature thermal
conductivity has the expression \cite{Gusynin}
\begin{equation}
\frac{\kappa}{T} \propto
\frac{\Gamma_{\mathrm{imp}}^{2}}{\Gamma_{\mathrm{imp}}^{2} + m^{2}}.
\end{equation}
It is no longer universal and depends explicitly on impurity
scattering rate as well as on gap $m$. Obviously, in the weak
impurity limit, $\Gamma_{\mathrm{imp}} \ll m$, the thermal
conductivity is rapidly suppressed from its universal value by $m$.
To see how $\kappa$ varies with magnetic field $H$, we need to know
the field dependence of gap $m(H)$. To this end, we solved the gap
equation (21) numerically \cite{note2} and presented the results for
$\alpha=4$ and $g=200$ in Fig.2. It is evident that the gap $m(H)$
is an increasing function of magnetic field $H$. Thus in the
symmetry broken phase, as magnetic field $H$ grows, the thermal
conductivity is suppressed and the system undergoes a phase
transition from thermal metal to thermal insulator at high field
limit, which is qualitatively in agreement with transport
experiments \cite{Hawthorn,Sun03}.

\begin{figure}[ht]
\begin{center}
\epsfxsize=0.40\textwidth \epsfbox{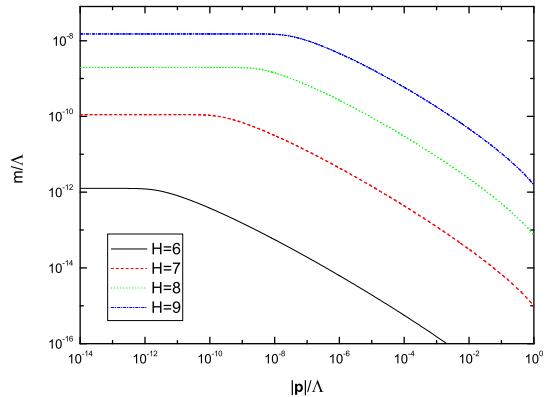}
\end{center}
\caption{The dependence of the generated gap on momentum and
magnetic field for $\alpha=4$ and $g=200$. $\Lambda$ is the momentum
cutoff.} \label{fig2}
\end{figure}

As revealed by transport experiments, the field induced reduction
of thermal conductivity occurs only in underdoped and optimally
doped cuprate superconductors \cite{Hawthorn, Sun03}. In the
overdoped region, the thermal conductivity is found to increase
with growing magnetic field $H$ \cite{Hawthorn, Sun03}. This can
be understood by assuming that the $d$-wave superconductivity
responses differently to external magnetic field in underdoped and
overdoped cuprates. On the underdoping side, the magnetic field
only reduces the superfluid density inside vortex cores but leaves
the $d$-wave energy gap essentially unchanged. Due to the
additional excitonic gap along nodal directions generated by
Coulomb interaction, the low energy nodal quasiparticles are
significantly suppressed, thus reducing the thermal conductivity.
However, on the overdoping side, the magnetic field destroys the
$d$-wave energy gap by directly breaking the Cooper pairs. As a
result, extended quasiparticles are excited from the condensate
and the thermal conductivity increases with growing field $H$.

\section{Summary and Discussion}

In summary, we proposed a mechanism to explain the field-induced
reduction of thermal conductivity in the mixed state of $d$-wave
cuprate superconductor. In this mechanism, a finite gap for nodal
fermions is generated by the strong Coulomb interaction between
gapless particle and hole excitations. Since the Coulomb
interaction is usually weak in the superconducting state, such gap
generation is possible only after the superfluid density is
reduced by strong external magnetic field. The excitonic gap
reduces the thermal conductivity at low temperature, so there is a
thermal metal-insulator transition driven by magnetic field.

There are several effects that might change the critical behavior of
chiral phase transition. For example, the long-range Coulomb
interaction can be screened by the finite zero-energy density of
states produced by disorder scattering and/or vortex scattering.
Such screening effect reduces the possibility of fermion gap
generation \cite{Liu08}. Only in clean superconductor and at
magnetic field much lower than the up critical field $H_{c2}$, this
effect can be ignored. On the other hand, the gap generation can
also be promoted by other mechanisms. If there are strong contact
interaction between nodal fermions, the possibility of pairing
instability is significantly enhanced due to the positive
contribution from contact interaction \cite{Liu08}. These competing
effects can be included into the above calculations along the steps
presented in Ref.\cite{Liu08}.

Throughout the present paper, the thermal fluctuation effect is
totally omitted. This effect actually plays at least three
important roles. First, the thermal fluctuation effectively
excites quasiparticles out of the condensate and hence reduces the
superfluid density rapidly. Secondly, these thermally excited
quasiparticles lead to screening of the Coulomb interaction. In
addition, the excitonic paring will surely be suppressed by
thermal fluctuations. These effects compete with each other,
making the situation rather complicated. This is why we simply
neglect the thermal effects and consider only nearly zero
temperature. To make an extension to finite temperatures, all
these three effects should be carefully analyzed.

Finally, we must admit that in the present work we utilized a number
of assumptions and approximations when studying the dynamical gap
generation for Dirac fermions. The results obtained in this paper
are only qualitatively reliable. In particular, we have not arrived
at a quantitative determination of the critical magnetic field
$H_{c}$. Generally speaking, $H_{c}$ must be a function of doping
concentration, temperature, and type of superconductor sample.
Unfortunately, it is difficult to quantitatively include any of
these effects. We wish the present work will be put on a firmer
theoretical ground in the future.

\section*{Acknowledgments}

We thank Wei Li for helpful discussions. This work was supported
by the NSF of China No.10674122.


\begin{thebibliography}{99}
\bibitem{note1}
The coupling $g$ is related to the interaction parameter $\lambda$
defined in graphene \cite{Liu08} by $g = 8\lambda$.
\bibitem{note2}
In a more rigorous treatment, the full solution of gap function
$m(\mathbf{p},H)$ should be obtained by sovling the closed integral
equations of gap function (21) and of superfluid density (16).
\end{thebibliography}
\end{document}